*processes*

MDPI

*Article*

# Model-based Stochastic Fault Detection and Diagnosis for Lithium-ion Batteries

Jeongeun Son, Yuncheng Du*

Department of Chemical & Biomolecular Engineering, Clarkson University, Potsdam NY 13676, USA
* Correspondence: ydu@clarkson.edu; Tel.: +1-315-268-2284

**Abstract:** Lithium-ion battery (Li-ion) is becoming the dominant energy storage solution in many applications such as hybrid electric and electric vehicles, due to its higher energy density and longer life cycle. For these applications, the battery should perform reliably and pose no safety threats. However, the performance of Li-ion batteries can be affected by abnormal thermal behaviors, defined as faults. It is essential to develop reliable thermal management system to accurately predict and monitor thermal behaviors of Li-ion battery. Using the first-principle models of batteries, this work presents a stochastic fault detection and diagnosis (FDD) algorithm to identify two particular faults in the Li-ion battery cells, using easily measured quantities such as temperatures.

Models of Li-ion battery are typically derived from the underlying physical phenomena. To make model tractable and useful, it is common to make simplifications during model development, which may consequently introduce mismatch between models and battery cells. Further, FDD algorithms can be affected by uncertainty, which may originate from either intrinsic time varying phenomena or model calibration with noisy data. A two-step FDD algorithm is developed in this work to correct model of Li-ion battery cells and to identify faulty operations from a normal operating condition. An iterative optimization problem is proposed to correct the model by incorporating the errors between measured quantities and model predictions, which is followed by an optimization-based FDD to provide a probabilistic description of the occurrence of possible faults, while taking the uncertainty into account. The two-step stochastic FDD algorithm in this work is shown to be efficient in terms of fault detection rate for both individual and simultaneous faults in Li-ion batteries, as compared to Monte Carlo (MC) simulations.

**Keywords:** fault detection and classification; uncertainty analysis; lithium ion battery; optimization; thermal management; polynomial chaos expansion

## 1. Introduction

Lithium-ion (Li-ion) batteries are widely used in many applications such as cell phones, electric and hybrid electric vehicles, since they exhibit higher energy density and have relatively longer life as compared to other batteries [1]. In these systems, Li-ion batteries must possess high reliability and pose no safety threats [2]. However, the thermal behavior can greatly affect the safety, durability, and performance of Li-ion batteries [3]. For example, fire and explosion caused by thermal runaway were reported [4]. Thus, reliable battery management systems are essential to mitigate negative effects (e.g. thermal runaway) and avoid catastrophic failures [5]. As a key component of the battery management system, fault detection and diagnosis play an important role in the management of Li-ion batteries [6].

Fault detection and diagnosis (FDD) methods generally can be classified into two major groups, i.e., first-principle model-based methods and data driven (or empirical) methods [7]. For the former, models describing physical mechanisms of the fault dynamics are oftentimes used, while historical data are typically collected for data driven methods to derive empirical models. Each of these approaches has its own advantage and drawback depending on specific problems. It is recognized that first-principle model-based methods exhibit better extrapolation ability, whereas data-driven methods are easier to design [8]. This work focuses on the use of the first-principle models for FDD, since these models provide fundamental understanding of the thermal physics of batteries [9].

Several first-principle thermal models have been previously developed for Li-ion batteries. For example, a three-dimensional thermal finite element model was developed to investigate the cell behavior under abnormal events such as overheating and external short circuits [10]. This model requires high computational capabilities, and its application is limited to stationary storage [11]. As compared to the three-dimensional models, one-dimensional model of Li-ion batteries, developed using the average lumped temperature of the cell, is viable for real-time applications and can enable online battery management [12]. However, such a model may fail to provide insights of the thermal (fault) dynamics due to its simplicity [13]. As a trade-off, a two-dimensional thermal model was developed, which can predict the core and the surface temperature of the Li-ion battery cells [3, 13]. Since the two-dimensional model can provide a better understanding of the thermal dynamics of battery cells, while maintaining the computational complexity, it is used in this work for the design of a stochastic FDD scheme.

Measurements of temperatures such as surface and core temperatures are often used for FDD in Li-ion batteries, but there is no direct measurement of the core temperature. To take the core temperature into account, estimation techniques are often required. In the literature, several estimation techniques have been developed. For example, adaptive observer based on lumped thermal model [14] and state observer using partial differential algebraic equations [15] were proposed to estimate the temperature. As compared to these estimation techniques, the real-time monitoring and diagnosis of faults in batteries are less explored. Although there have been several proposed works related to diagnostic algorithms for internal faults in Li-ion batteries [3, 16, 17], it is important to note that previously reported FDD works mostly investigated sensors or actuator fault detection problems [18, 19, 20].

In this work, we propose to estimate the core temperature and further use the estimation results to identify and classify two sets of faults. That is, faults that can introduce dynamic changes in core temperatures and faults that can affect the surface temperatures. The FDD scheme in this work can potentially provide more information about the thermal dynamics of batteries and enable internal thermal fault detection to improve the performance of Li-ion battery.

For FDD, the available algorithms compare the observed behavior to the corresponding model results, estimated from first-principle models [21]. When a fault is detectable, the FDD scheme will generate fault signatures, which in turn can be referred to an FDD scheme to identify the root cause of faults using a threshold [22]. However, the main restrictive factor for the first-principle model-based FDD is the model uncertainty [23]. The accuracy of fault detection algorithm can be affected by any uncertainty in model parameters. Such an uncertainty may result from intrinsic time varying phenomena or originate from model calibration with noisy measurements [24]. The uncertainty can be quantitatively approximated by calibration with experimental data, which include principles such as least squares errors or the Delphi method [25, 26].

The procedures that firstly quantify the uncertainty and then propagate the uncertainty onto the FDD scheme are typically omitted in previously reported works. This subsequently may lead to a loss of information about the effect of uncertainty on FDD performance. Recently, several techniques, such as adaptive observer [27, 28] and sliding mode observer [29], were developed for FDD in the presence of uncertainty. However, most of these methods cannot provide information such as the probability that a fault has occurred. In addition, since the faults in the batteries may happen in a stochastic fashion, the use of fixed thresholds to identify the root cause of faults may not be effective.

There are differences between the actual thermal dynamics of Li-ion batteries and fundamental models derived from physical phenomena. For example, to make models tractable and useful, it is common to make simplifications during the model development, which will introduce mismatch between the model and the Li-ion battery system of interest. Thus, the first principle model-based FDD scheme should be designed to compensate the mismatch. Specifically, a set of fixed model parameters may not be accurate enough for estimating the core temperature in the presence of model mismatch. Consequently, any inaccuracy in temperature estimation may potentially lead to low fault detection rate. To ensure the accuracy of FDD, it is essential to simultaneously calibrate the model parameters and adjust the FDD scheme. However, this is generally challenging due to the presence of uncertainty such as measurement noise and unknown model mismatch.

In this work, we propose to address these aforementioned limitations by developing an FDD scheme for Li-ion batteries described by a two-dimensional first-principle thermal dynamic model, for which both model parameters and faults are of stochastic nature. Specifically, the faults considered in this work such as the thermal runaway are stochastic perturbations superimposed on step changes in specific thermal dynamic parameter and electric current. The objective is to identify the changes in the mean values of thermal dynamic parameter and the current in the presence of the random perturbations, measurement noise, and model mismatch. As compared to other existing thermal diagnostic techniques, the main feature of the FDD scheme is the efficient quantification of the effect of stochastic changes in model parameters on fault detection, and the rapid propagation of the stochasticity onto the estimation of temperatures that are required for FDD.

Note that one possible way to propagate uncertainty in model parameters onto temperature estimates is the use of Monte Carlo (MC) simulations [30]. However, methods such as MC may be computationally demanding, since they often require a larger number of simulations in order to obtain accurate results. It is worth mentioning that although the calibration of an FDD scheme can be performed offline, but the online re-calibration of the model in the presence of model mismatch with MC as shown later in current work is computationally prohibitive. Recently, uncertainty propagation with generalized Polynomial Chaos (gPC) expansion has been studied in different modelling [31], optimization [32], and fault detection problems [24]. As compared to MC, the advantage of gPC is that it can propagate a complex probability distribution of uncertainty in model parameters onto model predictions rapidly and can analytically approximate the statistical moments of model predictions in a computationally efficient manner [31]. The improvement in computational time may facilitate its application in real-time model adjustment for improved FDD.

The FDD algorithm in this work is specifically targeted to identify and diagnose stochastic thermal faults consisting of uncertainty around a set of mean values of thermal properties in the presence of model mismatch. In summary, the contributions in this work include: (*i*) the use of an intrusive gPC model for stochastic FDD of Li-ion batteries by approximating the uncertainty in thermal dynamics with gPCs and by propagating the uncertainty directly onto temperatures that can be used for FDD; (*ii*) the identification and classification of a fault based on the probability information of temperatures other than a single point estimate or threshold; (*iii*) the formulation of an optimization to account for model mismatch and adjust the thermal dynamic models by incorporating the discrepancy between model predictions and measurements.

This paper is organized as follows. Section 2 presents the theoretical background and the principal methodologies in this work, including a two-dimensional thermal dynamic model, introduction of generalized polynomial chaos (gPC) expansion, and formulation of the stochastic fault detection and diagnosis (FDD) problem. The methodology for FDD and the formulation of an optimization for model correction to account for model mismatch is presented in Section 3. Analysis and discussion of the results are given in Section 4, followed by conclusions in Section 5.

## 2. Theoretical Backgrounds

### 2.1. Thermal Model of Lithium-ion Battery

The two-dimensional deterministic thermal dynamic model is used to describe a cylindrical Li-ion battery cell in this work [3, 13]. A schematic diagram of the Li-ion battery cell is shown in Figure 1. This model can provide information about the heat source of battery and estimate the core temperature based on measurements of surface temperature. The surface temperature $T_s$ and the core temperature $T_c$ can be defined as:

$$C_c \dot{T}_c = I^2 R_e + \frac{T_s - T_c}{R_c} \tag{1}$$

$$C_s \dot{T}_s = \frac{T_f - T_s}{R_u} - \frac{T_s - T_c}{R_c} \tag{2}$$

$$R_e = \beta_0 + \beta_1 SOC + \beta_2 T_c \tag{3}$$

where $I$ is the current, $T_f$ represents the surrounding air temperature, $R_e$ is the internal (or electrical) resistance, $R_c$ is the thermal resistance between the surface and core of the battery, $R_u$ denotes the convection resistance between the surface and the surrounding of the battery, $C_c$ and $C_s$ represent the heat capacity of the internal battery material and the surface battery material, respectively. The internal resistance $R_e$ is given in Equation (3) which consists of state of charge (*SOC*), core temperature $T_c$, and parameters $\beta_0$, $\beta_1$, $\beta_2$ that can be pre-estimated by an offline estimation scheme [3].

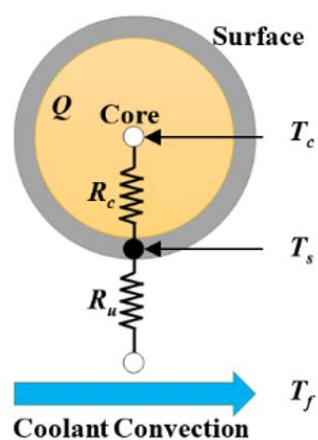

**Figure 1.** Schematic of thermal model of Li-ion battery cell.

For the Li-ion battery cell model given in Equations (1) and (2), model parameters including $R_e$ are generally assigned with constant values. A set of parameters used in the two-dimensional thermal dynamic model is given in Table 1 [33].

**Table 1.** Parameter Declaration for the Thermal Model of Li-ion Battery Cell.

| Model parameters | $C_c$ | $C_s$ | $R_e$ | $R_c$ | $R_u$ |
|---|---|---|---|---|---|
| Units | JK$^{-1}$ | JK$^{-1}$ | m$\Omega$ | KW$^{-1}$ | KW$^{-1}$ |
| Value | 268 | 18.8 | 10 | 2 | 1.5 |

It is important to note that the model of battery and the model parameters may involve uncertainty. For example, the thermal dynamics of a Li-ion battery cell can change with respect to time, which may be caused by factors such as the surrounding temperature and the state of charge. In addition, the estimates of model parameters can be affected by noisy data used for model calibration. These possible sources of uncertainty can be briefly categorized into three groups as follows.

1. *Observational uncertainty*: This includes measurement errors in experimental data such as the measurements of voltage, current, and surface temperatures.
2. *Parametric uncertainty*: This refers to uncertainty in parameters, which may originate from the observational uncertainty or result from lack of information. It may be advantageous to represent a model parameter, e.g., $R_e$ in Equation (1), as a random variable with a distribution other than a fixed value.

3. *Structural uncertainty*: This describes the differences between a model and the actual Li-ion battery system. For example, models in Equations (1) and (2) may not be an exact representation of the thermal dynamics of a Li-ion battery cell.

In this current work, we focus on the development of FDD algorithms in the presences of these uncertainties. Specifically, the conduction resistance $R_c$ in Equations (1) and (2) is considered as an uncertain parameter and changes in $R_c$ are defined as stochastic faults. The conduction resistance $R_c$ is often used to incorporate conduction and thermal resistance across materials with compact and inhomogeneous properties. It is difficult to accurately estimate the exact parameter value of $R_c$, since the rolled electrodes consist of the cathode, anode, separator, and current collectors, which may complicate the parameter estimation and reduce the estimation accuracy [14]. Any variations in $R_c$, may significantly affect the performance of battery. In addition, it is assumed that the current $I$ in Equation (1) is the second uncertainty in this work, since the internal state of the battery can be affected by current [34]. For example, as previously reported [14], current variations may lead to the fluctuation in temperatures of the battery. Furthermore, the electric current of battery can be time-varying in practice and can be corrupted by measurement errors, thus the exact value of current can be an unknown prior.

Since the convection resistance $R_u$ is related to the surrounding coolant flowrate [35], which is oftentimes tightly controlled to maintain a consistent battery temperature, thus $R_u$ is assumed to be a constant other than a parametric uncertainty. For the internal resistance $R_e$ in Equation (1), it can be affected by various conditions such as the state of charge of battery, temperature, and drive cycle [14, 36, 37] leading to the changes in model predictions such as temperature. However, this thermal parameter in Li-ion battery has been investigated by many researchers and is well formulated with the state of charge and temperatures as shown in Equation (3) [3, 14, 38]. For example, it can be estimated offline with experimental data or determined online with SOC estimation based on an equivalent circuit model (ECM) [38]. In this work, it is assumed that $R_e$ is constant rather than time-varying parameter and it is not considered as a parametric uncertainty for simplicity. However, the proposed uncertainty propagation and diagnostic scheme can be extended to $R_u$ and $R_e$ according to their intrinsic properties when there is evidence to support significant variation in $R_u$ and $R_e$.

In this work, sudden changes of temperatures in Li-ion battery caused by the current $I$ and resistance $R_c$ will be diagnosed and classified by the proposed method. Additionally, to introduce structural uncertainty, it is assumed that the exact statistical moments of uncertainties, such as the actual mean value of $R_c$ is unknown to the modelers, which will be corrected by incorporating the differences between model predictions and measurement of temperatures. Further, it should be noted that only the surface temperature of battery can be directly measured, thus the estimations of the core temperatures will be used in the model correction.

*2.2. Generalized Polynomial Choas Expansion*

The generalized polynomial chaos (gPC) expansion approximates a random variable with an arbitrary probability density function (PDF) of another random variable (e.g., $\xi$) with a known prior distribution. For brevity, suppose that the battery thermal models in Equations (1) and (2) can be described by a set of ordinary differential equations (ODEs) as:

$$\dot{x} = f(t, x, u, p) \tag{4}$$

where the vector $x=\{x_j\}$ (j=1, 2, …, n) represents the core and the surface temperatures, i.e., $T_c$ and $T_s$, with initial values $x_0$ at $t=0$, $u$ is deterministic parameters, i.e., fixed constant values, while $p$ is a vector of uncertainties, i.e., $I$ and $R_c$ in this work, which will be approximated with PDFs. To evaluate the effect of uncertainty on temperatures, a key step is to approximate each parameter in $p=\{p_i\}$ (i=1,2, .., $n_p$) as a function of a set of independent random variable $\xi=\{\xi_i\}$ as:

$$p_i = p_i(\xi_i) \tag{5}$$

where $\xi_i$ denotes the $i^{th}$ independent random variable following a standard PDF [31]. Based on the definition of gPC expansion, each parametric uncertainty $\{p_i\}$ and the model predictions $x$ can be defined using the orthogonal polynomial basis functions $\{\phi_k(\xi)\}$ as:

$$p_i(\xi_i) = \sum_{k=0}^{\infty} \hat{p}_{i,k} \phi_k(\xi_i) \tag{6}$$

$$x_j(t, \xi) = \sum_{k'=0}^{\infty} \hat{x}_{j,k'}(t) \varphi_{k'}(\xi) \tag{7}$$

where $\{\hat{p}_{i,k}\}$ denote the gPC coefficients of the $i^{th}$ parametric uncertainty, $\{\hat{x}_{j,k'}\}$ are the gPC coefficients of the $j^{th}$ model predictions at time instant $t$, and $\{\varphi_{k'}(\xi)\}$ are the orthogonal polynomial basis functions of random variables $\xi$ [31]. When the PDFs of $p$ are a given prior, a set of coefficients $\{\hat{p}_{i,k}\}$ in Equation (6) can be determined such that $p_i(\xi_i)$ follows *a prior* known distribution. Otherwise, optimization techniques can be used to estimate $\{\hat{p}_{i,k}\}$. As compared to $p$, the gPC coefficients of $x$ are unknown and have to be calculated. To calculate $\{\hat{x}_{j,k'}\}$, Equations (6) and (7) are firstly substituted into Equation (4), which is followed by applying a Galerkin projection and by projecting Equation (4) onto each of the polynomial chaos basis function $\{\varphi_{k'}(\xi)\}$ as:

$$\langle \dot{x}_j(t,\xi), \varphi_{k'}(\xi) \rangle = \langle f(t, x_j(t,\xi), \mathbf{u}, \mathbf{p}(\xi)), \varphi_{k'}(\xi) \rangle \tag{8}$$

For practical application, truncation, i.e., a finite number of terms, is often used other than infinite terms in Equations (6) and (7). For example, the total number of approximation terms (i.e., $Q$) that can be used for $\{x_j\}$ in Equation (7) can be calculated as:

$$Q = ((n_p + q)! / (n_p! q!)) - 1 \tag{9}$$

where $q$ is the number of terms that is necessary to approximate an arbitrary uncertainty with a prior known PDF in Equation (6), and $n_p$ is the total number of parametric uncertainties in $p$. As seen in Equation (9), the number of terms required for the gPC approximation of $x=\{x_j\}$ depends on the order of polynomial $q$ and/or the number of unknown parametric uncertainty $n_p$.

The inner product between any two vectors in Equation (8) can be calculated as [31]:

$$\langle \psi(\xi), \psi'(\xi) \rangle = \int \psi(\xi) \psi'(\xi) W(\xi) d\xi \tag{10}$$

where the integral is calculated over the entire domain defined by random variables $\xi$ in the Wiener-Askey framework, $W(\xi)$ is the PDF of $\xi$ that is defined as a weighting function in gPC theory. For example, Hermite polynomial basis functions can be used for normal distributions [31]. Using gPC coefficients of model predictions $x$ in Equation (7), the statistical moments of $x$ at a given time $t$ can be quickly estimated as follows:

$$E\left(x_j(t)\right) = E\left[\sum_{k'=0}^{Q} \hat{x}_{j,k'}(t) \varphi_{k'}\right] = \hat{x}_{j,0}(t) E(\varphi_0) + \sum_{k'=1}^{Q} E[\varphi_i] = \hat{x}_{j,0}(t) \tag{11}$$

$$Var\left(x_j(t)\right) = E\left[\left(x_j(t) - E[x_j(t)]\right)^2\right] = E\left[\left(\sum_{k'=0}^{Q} \hat{x}_{j,k'}(t) \varphi_{k'} - \hat{x}_{j,k'=0}(t)\right)^2\right]$$
$$= E\left[\left(\sum_{k'=1}^{Q} \hat{x}_{j,k'}(t) \varphi_{k'}\right)^2\right] = \sum_{k'=1}^{Q} \hat{x}_{j,k'}(t)^2 E(\varphi_{k'}^2) \tag{12}$$

In addition, the PDF of model predictions $x$ can be estimated by sampling from the PDF of $\xi$ and by substituting samples into the gPC expressions of $x$ in Equation (7). The calculation of statistical moments with the analytical formulae in Equations (11) and (12) and the rapidly approximation of the PDF of $x$ are the main rationale of using the gPC in this current work, since it can reduce the computational burden involved in the model correction in the presence of structural and parametric uncertainty. Note that the FDD procedure in this work consists of the inverse of the procedures summarized above, i.e., the identification of the PDFs (e.g., mean values) of parametric uncertainty using the measurements and model predictions of $x$. The details about the FDD will be discussed in Section 3.

*2.3. Formulation of FDD problem*

The faults considered in this work consist of stochastic perturbations superimposed on a particular set of mean values of these two aforementioned uncertainties, *i.e.*, current *I* and conduction resistance $R_c$. For example, Figure 2 shows a possible fault profile (Figure 2 (a)) and the resulting noise-free temperature responses (Figure 2 (b)). For clarity, two mean values of each faults in Figure 2 are presented. As can be seen, any changes in the mean values of faults can induce variations in temperatures. The objective is to use the measurements of temperature to identify the step changes between different mean values of current (*I*) and thermal resistance $R_c$.

A mathematical description of stochastic faults is defined as:

$$p_i = \bar{p}_i + \Delta p_i (i = 1, \dots, n_p) \tag{13}$$

where $p_i \in p$ (i=1,2, ..., $n_p$), $\{\bar{p}_i\}$ denotes a set of mean values, and $\{\Delta p_i\}$ represents the variation around each mean value of the $i^{th}$ uncertainty. For example, the solid bold lines (blue and red) in Figure 2 (a) are the mean values of current (*I*) and thermal resistance $R_c$, while the purple and green lines are the perturbations around each of the mean values. It is assumed in this work that the statistical moment of $\{\Delta p_i\}$ is *time-invariant* for simplicity and can be estimated with offline model calibration algorithms. In addition, the total number of possible mean values of $p_i$ can be experimentally inferred from the constancy of measured quantities such as the surface temperature as shown in Figure 2 (b), but the exact mean values can be unknown to the modelers.

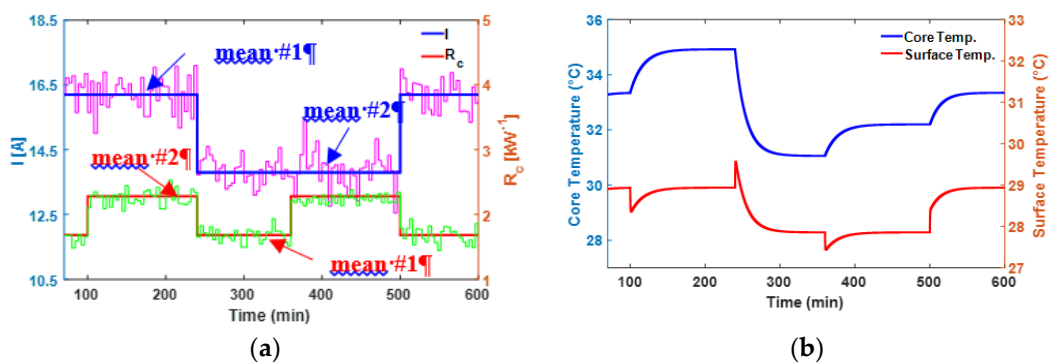

(a) (b)

**Figure 2.** Profiles of faults (a) and the corresponding noise-free temperature (b). Note that the purple and green lines in (a) represent the perturbations around the mean values of possible faults, and noise free measurements of temperatures are used in (b) for clarity.

As seen in Figure 2 (b), the core temperature is higher, when the mean values of *I* and $R_c$ are larger. Since any significant changes in the core temperatures are harmful and may cause catastrophic failures in Li-ion batteries [4], thus the smaller mean values of *I* and $R_c$ are used to represent the normal operating mode of Li-ion battery in this work, while the larger mean value in either *I* and $R_c$ is used to represent the faulty operating modes. Thus, the objective is to identify the mean value (or mean value changes) of *I* and $R_c$ in the presence of uncertainty.

To summarize, two types of faults are considered. (*i*) *Fault 1*: Current fault (*I*), representing the switch between two mean values of *I*, which can affect the core temperature dynamics and further induce thermal runaway faults. (*ii*) *Fault 2*: Thermal resistance fault ($R_c$), representing a significant deviation in the mean value of thermal resistance $R_c$, which may result from battery aging and can affect both core and temperatures. Based on the definition of the faults, the setting of normal and faulty operating modes in this work is given in Table 2, respectively.

**Table 2.** Faults Definition and Description.

| Modes | Description | Type |
|---|---|---|
| Normal | $I = \bar{I}^1$, $R_c = \bar{R}_c^1$ | No fault |
| Faulty 1 | $I = \bar{I}^2$, $R_c = \bar{R}_c^1$ | Individual fault |
| Faulty 2 | $I = \bar{I}^1$, $R_c = \bar{R}_c^2$ | Individual fault |
| Faulty 3 | $I = \bar{I}^2$, $R_c = \bar{R}_c^2$ | Simultaneous faults |

## 3. Methodology of Fault Detection and Diagnosis

The objective of the FDD algorithm is to identify a change in the mean values of $I$ and $R_c$ and classify an operating condition as a normal or faulty mode described in Table 2, using measurements of temperatures. A Joint Confidence Region (JCR) based FDD algorithm is first presented in Section 3.1, which is followed by an optimization-based model correction method in Section 3.2 for improved FDD in the presence of model mismatch.

### 3.1. Fault Detection Algorithm using JCR Profiles

In Section 2, the propagation of uncertainty onto model predictions was discussed, from which the PDF profile of each model prediction can be approximated using the gPC models. The main idea of the FDD algorithm in this work is to solve the inverse problem, *i.e.*, to identify the mean values of uncertainty with gPC models. The FDD method consists of three steps. (*a*) The stochasticity in faults (i.e., $I$ and $R_c$) is propagated onto model predictions, thus producing a family of gPC models of the core and surface temperatures around each mean value of faults considered in this work. (*b*) Since two uncertainties (faults) are studied, a set of joint confidence region (JCR) profiles of the core and surface temperatures is used to infer the possible mean values or any changes in mean values of faults. The generation of the JCR will be discussed later, which predicts the probability that a pair of measurements belongs to a particular JCR. (*c*) Because of the measurement noise and the overlaps among JCRs, the JCR-based FDD may provide lower fault detection rate. Thus, a gPC model-based minimum distance optimization is developed to improve the FDD performance.

*Step a*

The formulation of the gPC models for the core and surface temperatures follows the procedures as outlined in Section 2. It is assumed that the stochastic perturbations in faults $I$ and $R_c$ are independent stochastic events, thus a two-dimensional random space is used, *i.e.*, $\xi = \{\xi_1, \xi_2\}$. Consequently, the predictions of temperatures obtained from Equation (7) are functions of $\xi = \{\xi_1, \xi_2\}$, *i.e.*, any changes in faults can affect both core and surface temperatures.

*Step b*

Since two faults are studied, JCR profiles of core and surface temperatures are used to infer mean value changes in faults $I$ and $R_c$. Figure 3 shows a schematic of generated JCRs from gPC models. The generation of JCRs proceeds as follows.

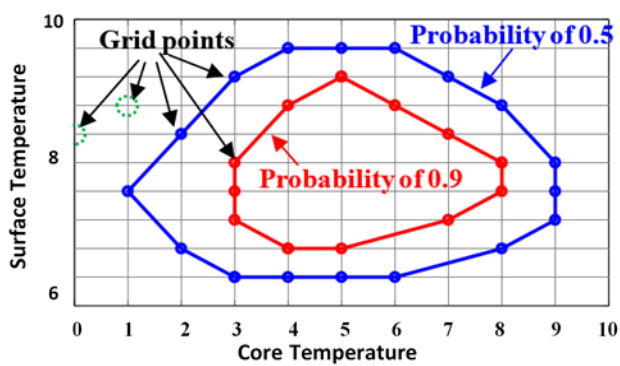

**Figure 3.** Schematic of generated JCRs with different probabilities. Note that the units of temperatures in this work is Celsius degree (°C).

(*i*) In the case of stochastic perturbations in both *I* and $R_c$, the maximum variations of core and surface temperatures are first estimated. (*ii*) A two-dimensional discrete domain made of combinations of core and surface temperature values can be generated based on the temperature estimations in *Step i*. (*iii*) Random samples of $\xi_1$ and $\xi_2$ are substituted into the gPC models of the core and surface temperatures as defined in Equation (7), which can provide the temperatures values. (*iv*) Each pair of the core and surface temperatures is assigned to a particular grid generated in *Step ii*, and the total number of temperature pairs can be calculated when all the samples from *Step iii* have been assigned. (*v*) The probability at each discrete grid is calculated as the ratio between the number of temperature pairs at a particular grid point and the total number of temperature pairs (i.e., the number of combination of $\xi_1$ and $\xi_2$ that are used in *Step iii*). (*vi*) A JCR can be generated by connecting discrete grid points with the same probability (see Figure 3).

*Step c*

Following the procedures above, a family of JCR profiles can be generated for each pair of mean values of *I* and $R_c$, as shown in Table 2, which can be used for FDD. However, as seen in Figure 4 (a), the JCRs used to infer faults can be misleading, when a pair of measurements (red star) is found to be in the overlap of JCRs. In addition, the measurements may lay outside of JCR profiles due to the measurement noise, as shown in Figure 4 (b). Thus, a gPC model-based minimum distance criterion is used to improve the FDD performance, which is explained below.

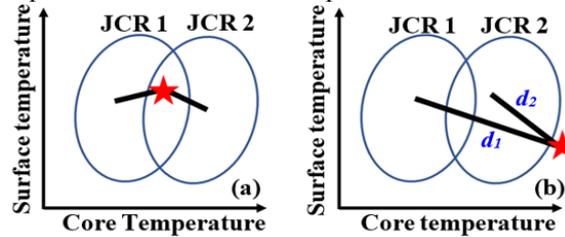

**Figure 4.** Visual interpretation of FDD algorithm using JCRs. Note that (a) represents that a pair of measurements can be found in the overlap of the JCRs, and (b) represent that a pair of measurements can be found outside the JCRs due to measurement noise. In addition, $d_1$ and $d_2$ in (b) represent the distance between the measurements and the centers of JCRs, which can be used for FDD with a minimum distance criterion as defined in Equations (14) and (15).

As seen in Equation (7), the gPC models of the core and surface temperatures are functions of random variables $\xi = \{\xi_1, \xi_2\}$, which can provide the statistical information of temperatures resulting from stochasticity in faults *I* and $R_c$. The combination of gPC models of core and surface temperatures can provide the mathematical description of JCRs. When a pair of temperatures is available, e.g., red star in Figure 4, it is possible to calculate the distance between a pair of temperatures and the center of a JCR. For a prescribed confidence region (or specific probability), the shortest distance between the measurements and a specific JCR can then be used to infer the mean values of faults. For example, as seen in Figure 4 (b), the distance $d_2$ is smaller than $d_1$, thus indicating that the mean values of faults, used to generate JCR-2, are the most probable operating mode. To analytically decide the Euclidean distance between a pair of measurements and a JCR, an optimization problem is developed as:

$$\min_{\lambda} J_i = (T_{c,i} - T_{c,p})^2 + (T_{s,i} - T_{s,p})^2 \tag{14}$$

$$Operating\ mode: M_{FCR} = arg\ min\{J_i\} \tag{15}$$

where *i* is the total number of combination of mean values of faults *I* and $R_c$ as shown in Table 2, $T_{c,i}$, and $T_{s,i}$ are the gPC models for a particular set of mean value *I* and $R_c$, which are functions of $\xi$ given in Equation (7), $T_{c,p}$, and $T_{s,p}$ are core and surface temperatures that are used for FDD. Note that $M_{FCR}$ in Equation (15) is the identified operating mode defined in Table 2 based on the minimum distance criterion. It should be noted that there is no direct measurement of core temperatures of battery, thus models, i.e., Equations (1) and (2), are used to estimate the core temperature with the measurement of surface temperature. The decision variable $\lambda$ is a vector of random samples of $\xi = \{\xi_1, \xi_2\}$ from the sample domain defined by the three-sigma rules [39]. This optimization problem in Equation (14) will be performed for each pair of core and surface temperature measurements and combination of mean values of faults *I* and *R_c* that are defined in Table 2. Then, the minimum distance as defined in Equation (15) can be used to identify an operating mode as defined in Table 2.

*3.2. Optimization-based Model Correction*

The FDD algorithm in Section 3.1 assumes that the exact statistical moments of *I* and *R_c* are given priors, which can be propagated onto the temperatures to formulate JCR profiles of temperatures. However, it cannot account for the discrepancy between the model and the actual thermal dynamics of Li-ion battery. For example, model calibration with noisy data can introduce model uncertainty. Further, model assumptions and simplifications are often made to make model tractable, which may result in structural uncertainty. To account for uncertainty (and/or mismatch) between the model and the actual battery cells, we propose to correct the model by incorporating the error between model predictions and available measurements. The correction criterion is formulated as follows:

$$\dot{\tilde{x}} = f(t, \tilde{x}, u, p) + \mu(\hat{x} - \tilde{x}) \tag{16}$$

where $\mu = \{\mu_j\}$ (j=1, 2, ..., n) is a vector of correction gains, $\tilde{x}$ is model predictions, and $\hat{x}$ is measurements of temperatures. To implement Equation (16), it is assumed that measurements of surface temperature are available, and the core temperature can be estimated with the model that is being corrected. It is also assumed that the exact statistical information such as mean value of the uncertainty is not available for the user, in order to represent a model involving model mismatch. Such a difference will be compensated using correction gains $\mu$ in Equation (16).

To calculate the correction gains, a set of measurements inside a sliding time window will be used in this work. A schematic of the sliding time moving window is shown in Figure 5, where *L* represents the size of the moving window and *M* is the moving rate, i.e., *L* determines a total number of required temperatures and *M* decides the overlap between the windows. A smaller window size can be less accurate and may be time consuming, but it can be sensitive as it would better capture the thermal dynamics of battery. A larger window size can reduce the computational burden, but it may lead to a coarse estimation. The moving rate decides the number of measurements changed at a time. For example, when 1 is used for *M*, which means that the one measurement is changed at a time, *i.e.*, the first measurement in *L* will be removed and one new measurement will be appended to *L*. When *M* is larger, it may produce poor model correction result, while it will increase the computational load when *M* is smaller. The choice of *L* and *M* is problem specific and requires a trade-off, which can be determined with insights of the dynamic natures of batteries.

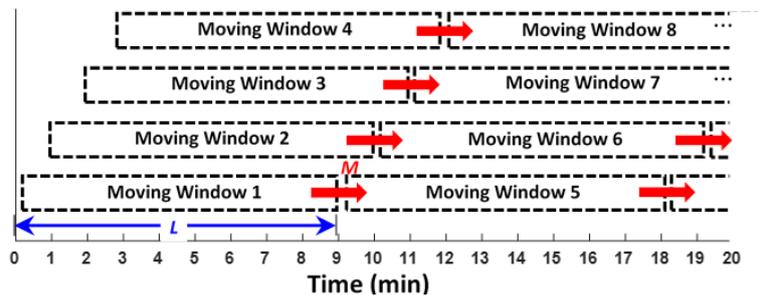

**Figure 5.** Schematic of sliding time moving windows for model correction

For a sliding time moving window with temperature measurements, the correction gains $\mu$ can be optimized with an optimization as:

$$\min_{\lambda=\mu} J = \sum_{i=1}^{L}(T_{c,i} - T_{c,p})^2 + \sum_{i=1}^{L}(T_{s,i} - T_{s,p})^2 \tag{17}$$

where $T_{c,i}$, and $T_{s,i}$ are gPC model predictions of core and surface temperatures obtained from Equation (16), $T_{c,p}$ and $T_{s,p}$ denotes the temperatures inside moving windows that are used for model correction. Note that core temperatures are estimated from the deterministic model that being corrected based on the measurements of surface temperatures. The decision variable $\lambda$ in Equation (17) is the correction gain that can be recursively updated with moving time windows. It will be shown in the results section that the model correction can be executed at each time interval in a real-time fashion, and the fault detection results can be greatly improved with the recursively-updated gPC model.

*3.3. Summary of FDD Algorithm*

An overview of the proposed model correction and FDD is shown in Figure 6. In summary, the algorithm proceeds as follows.

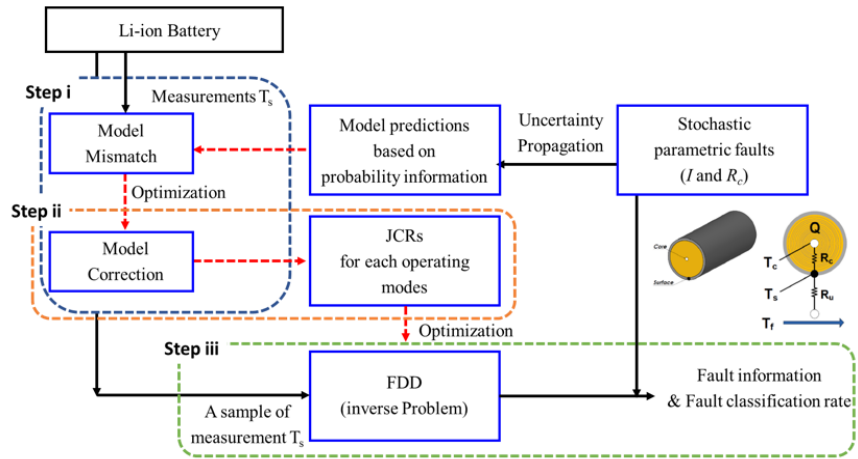

**Figure 6.** Overview of the proposed FDD algorithm

***Step i*** – Collect measurements of surface temperatures as a training set when the battery is operated at normal and faulty operating modes, described in Table 2. Using the optimization defined as Equation (17), the models of Li-ion battery cells can be corrected around each pair of the mean values of *I* and *Rc*. Note that measurements of temperatures for faults can be obtained from either historical database or designed experiments.

***Step ii*** – Using the corrected models, the JCR profiles of the core and surface temperatures for each operating mode can be generated following the procedures described in Section 3.1.

***Step iii*** – When a sample of surface temperature is available, the core temperature will be firstly estimated, and the minimum distance can be calculated with Equations (14) and (15), which can be used to infer a particular set of mean values of *I* and *Rc*.

To evaluate the performance of the proposed FDD approach, the fault classification rate ($r_{FCR}$) defined as below is used:

$$r_{FCR} = \frac{n_{id}}{n_{total}} \qquad (18)$$

where $n_{total}$ represents the total number of testing samples used for algorithm verification, and $n_{id}$ is the number of samples that have been correctly identified and classified.

## 4. Results and Discussion

*4.1. Uncertainty Propagation and Model Predictions*

The FDD algorithm is applied to the Li-ion battery cells as explained in Section 2.1. For clarity, two mean values of fault *I* and *Rc* are considered, respectively. For current fault, *I*, these mean values are $\bar{I}^1 = 16.2$ A and $\bar{I}^2 = 13.8$ A, respectively. It is assumed that the stochastic perturbations in *I* around each of these mean values follow a normal distribution with a mean of zero and a standard deviation of 0.45 A. For the conduction resistance *Rc*, two mean values are $\bar{R}_c^1 = 1.68$ KW$^{-1}$ and $\bar{R}_c^2 = 2.28$ KW$^{-1}$, respectively. In addition, the random variations around each mean value are normally distributed, which has a mean value of zero and a standard deviation of 0.066 KW$^{-1}$, i.e., a 5% variation with respect to the average of two mean values. Since the perturbations around the mean values follow a normal distribution, Hermite polynomial basis functions are used for gPC models in this work. It is important to note that for arbitrary distributions, the polynomial basis functions from the Askey-Wiener scheme other than Hermite polynomial basis functions can be used to improve the convergence of the gPC approximation in Equation (6) [31].

Following uncertainty propagation procedures described in Section 2.2, Figure 7 shows the mean of temperatures and the corresponding variance around the mean values at each time interval, when the battery is operated at the normal mode. Since two sources of uncertainty are studied (i.e., $n_p$ = 2 in Equation (9)) and two terms can be used to approximate a normally distributed $I$ or $R_c$ (i.e., $p$ = 1), six terms are required to approximate each temperature (i.e., $Q$ = 5 in Equation (9)). The gPC coefficients of the temperatures can be solved by substituting the gPC models of uncertainties and temperatures into the Li-ion battery model (Equations (1) and (2)), which can then be solved by a Galerkin projection as explained in Section 2.2. This will produce a set of coupled equations to describe the stochastic thermal dynamics of Li-ion battery cells. The resulting gPC models of the core and the surface temperatures are given by Equations (A1) to (A12) in the Appendix A for brevity.

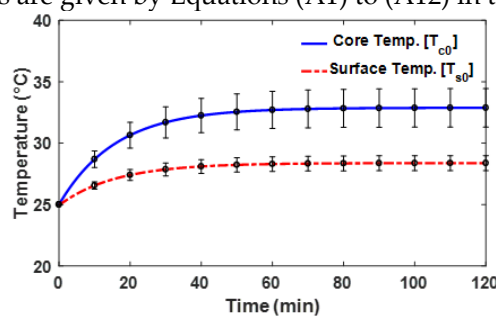

**Figure 7.** Uncertainty propagation in the lumped thermal models of Li-ion battery cell at the normal operating mode, (mean value of temperatures and the variance at a few particular time intervals).

As seen in Figure 7, $T_{c0}$ and $T_{s0}$ represent the mean values of core and surface temperatures, and the *bar-plots* represent the variances around the mean values which can be calculated from the higher order gPC coefficients, using Equation (12) in Section 2.2. Additionally, it was found that the core temperature can be significantly affected by variations in $I$ and $R_c$, as compared to the surface temperature, i.e., a larger variance as seen in Figure 7.

*4.2. FDD using JCR Profiles and Computational Efficiency*

Based on the gPC model developed with each pair of the mean values of $I$ and $R_c$, a family of JCRs can be generated following the procedures as explained in Section 3. Figure 8 shows the JCRs for a set of specific confidence regions, where 1000 pairs of temperatures samples are used. Based on the JCRs profile, the mean values of $I$ and $R_c$ can be inferred by solving the optimization problem defined in Equations (14) and (15) for a pair of temperatures. Taking a pair of temperatures as given in Figure 8 (the star) as an example, it can be concluded that the battery system is operated around the second set of mean values of $I$ and $R_c$, since the distance between the given samples of temperatures and JCR-2 is minimal. It should be noted that the JCR profiles not only can distinguish a specific faulty operating mode from the normal operation, but also provide probability information of being in a particular operating mode.

In addition, comparison studies were conducted to compare the gPC-based FDD with Monte Carlo (MC) simulations-based method. For MC, a similar optimization problem as done for the gPC is defined as:

$$\min_{\lambda'} J = \sum_{j=1}^{N}(T_c^j - T_{c,p})^2 + \sum_{j=1}^{N}(T_s^j - T_{c,p})^2 \tag{19}$$

where $\lambda'$ is the decision variables, i.e., the mean and the standard deviation of $I$ and $R_c$ that have to be determined with respect to a given pair of measurements of temperature, i.e., $T_{c,p}$, and $T_{s,p}$. Also, $N$ is the total number of samples used in the MC simulations in each iteration of the optimization, $T_c^j$ and $T_s^j$ are a particular set of core and surface temperatures simulated with respect to the decision variables. When the optimization of Equation (19) is finished, the optimization results $\lambda'$ are compared with mean values defined in Table 2 based on a minimum distance criterion, which can identify a corresponding operating mode.

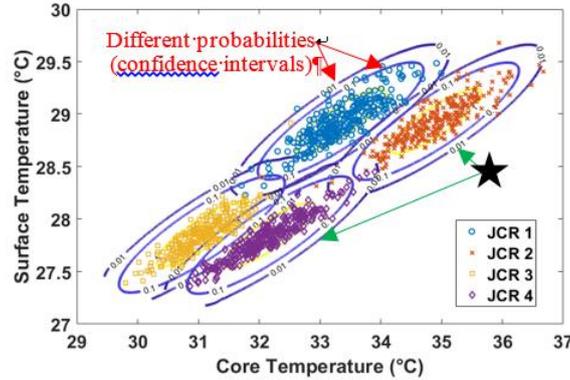

**Figure 8.** JCRs generated with a set of specific mean values of *I* and *R*$_c$, which are summarized in Table A1 in *Appendix B*. (*i*) JCR 1: 16.2 and 1.68 for *I* and *R*$_c$; (*ii*) JCR 2: 16.2 and 2.28 for *I* and *R*$_c$; (*iii*) JCR 3: 13.8 and 1.68 for *I* and *R*$_c$; (*iv*) JCR 4: 13.8 and 2.28 for *I* and *R*$_c$.

For the gPC-based FDD, it was found that the optimization problem described in Equations (14) and (15) can be finished within on average 5 seconds. However, for the MC-based method, the calculation of the mean values of *I* and *R*$_c$ on average requires approximately 321 seconds, when 100 pairs of samples of *I* and *R*$_c$ were used to simulate $T_c^j$ and $T_s^j$ in each optimization iteration. This clearly shows that the computational efficacy of gPC, as compared with MC. In addition, it was found that MC with 100 samples cannot provide as accurate results as gPC. For example, it was found that the fault classification rate *r*$_{FCR}$ of gPC and MC is ~0.94 and ~0.75, respectively. To improve the FDD performance, a larger number of samples are required in each iteration of the optimization with MC. However, this may significantly increase the computational burden. Especially, for the real-time model correction that will be discussed in next section, it can be computationally prohibitive with MC. A summary of the comparison between gPC and MC is given in *Appendix C*.

*4.3. FDD Results using JCRs in combination with Model Correction*

In previous case study, it is assumed that models of battery are accurate, and JCR profiles are used for FDD. In this section, the JCR profiles-based FDD algorithm is integrated with a model correction procedure to deal with FDD problem in the presence of model mismatch. For clarity, it is assumed that the exact mean values of *I* and *R*$_c$ for each operating modes (JCRs) are unknown to the modeler, thus a set of correction gains will be used to compensate the effect of model mismatch on FDD. Since the exact mean values of faults are unknown, the mean values in the gPC models of core and surface temperature are corrected using model predictions and measurements collected at each time interval inside the time moving windows, which can be described as:

$$\frac{dT_{c0}}{dt} = \frac{1}{C_c}\left(I_0^2 R_e + I_1^2 R_e + \frac{1}{R_{c0}}\left((T_{s0} - T_{c0})A + (T_{s2} - T_{c2})B + (T_{s4} - T_{c4})C\right)\right) + \mu_1(T_{c0} - \hat{T_c}) \quad (20)$$

$$\frac{dT_{s0}}{dt} = \frac{1}{C_s}\left(\frac{1}{R_u}(T_f - T_{s0}) - \frac{1}{R_{c0}}\left((T_{s0} - T_{c0})A + (T_{s2} - T_{c2})B + (T_{s4} - T_{c4})C\right)\right) + \mu_2(T_{s0} - \hat{T_s}) \quad (21)$$

where $T_{c0}$ and $T_{s0}$ are the first coefficients (i.e., mean values) in gPC models of the core and surface temperatures, $I_0$ and $R_{c0}$ are the gPC coefficients in Equation (6) used to approximate the mean values of *I* and *R*$_c$, $\hat{T_c}$ and $\hat{T_s}$ are the measurements of temperatures. Note that $\mu_1$ and $\mu_2$ are correction gains which will be recursively optimized with the optimization defined in Equation (17), $T_{s2}$, $T_{c2}$, $T_{s4}$, and $T_{c4}$ are higher order gPC coefficients of core and surface temperatures, which can be determined with gPC models as given in *Appendix A*. In addition, *A*, *B*, and *C* are constants calculated using gPC models with the Galerkin projection. For illustration, Figure 9 shows the model correction results of $\mu_1$ and $\mu_2$, when the system is operated at different operating modes as defined in Tables 2 and B1 in *Appendix B*. To introduce the model mismatch, a ±10% change was randomly added to these mean values given Table B1.

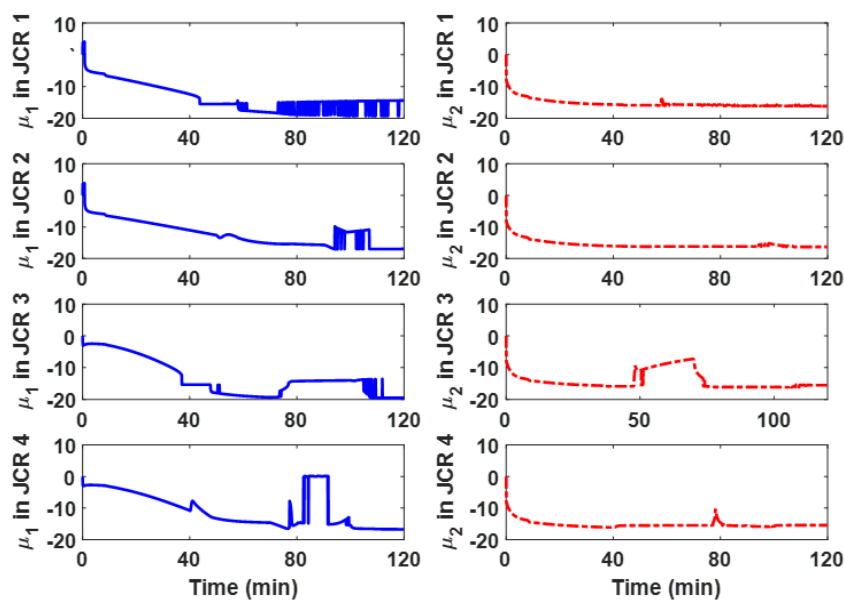

**Figure 9.** Correction gains $\mu_1$ and $\mu_2$ for different operating modes, where solid lines (blue) are the correction gain used for core temperatures and the dash-dotted line (red) are the results of surface temperature.

For different JCR profiles, the first column in Figure 9 represents the correction gains of the core temperature calculated at each time instant, whereas the second column is the correction gain of the surface temperature. As can be seen in Figure 9, the profiles of correction gains $\mu_1$ and $\mu_2$ fluctuate within a certain range when the optimization of Equation (17) was executed, and eventually reached a plateau. For example, the correction gain of the core temperature, i.e., $\mu_1$, varies significantly when the optimization was initially executed, e.g., 0 to ~80 minutes. In contrast, the changes in correction gains appear to be smaller after approximately 80 minutes of simulations. It is important to note that the perturbations in correction gains may either result from measurement noises or stochasticity in current *I* and conduction resistance $R_c$. In addition, it was found that correction gain $\mu_2$ of the surface temperature stabilizes faster than the correction gain of core temperature $\mu_1$. This is due to the fact that random variations in *I* and $R_c$ can significantly affect core temperatures as previously discussed in section 4.1 (see Figure 7). Note that the size of moving time window (*L*) is set to 80 for simulations as shown in Figure 9, i.e., 80 measurements were used to optimize the correction gains at each time instant. The moving rate *M* is set to 1 in this case study. In addition, random noise was added to the surface temperatures, which was further used to estimate core temperatures for optimization as defined in Equation (17).

Using these correction gains and the gPC coefficients, the distributions of the core and surface temperatures as each time interval can be rapidly estimated. For example, Figure 10 shows the simulation results of temperatures for the normal operation. Based on the corrected gPC models and the distributions of temperatures, a set of JCR profiles can be formulated and used for FDD following the steps as explained in Section 3.1.

To evaluate the efficiency of the correction and its effect on FDD, two case scenarios were investigated. For the first one, JCR profiles generated with the inaccurate mean values of *I* and $R_c$ were used, whereas the correction algorithm was combined with the JCR-based FDD in the second case scenario. Table 3 shows the results of FDD for both case studies.

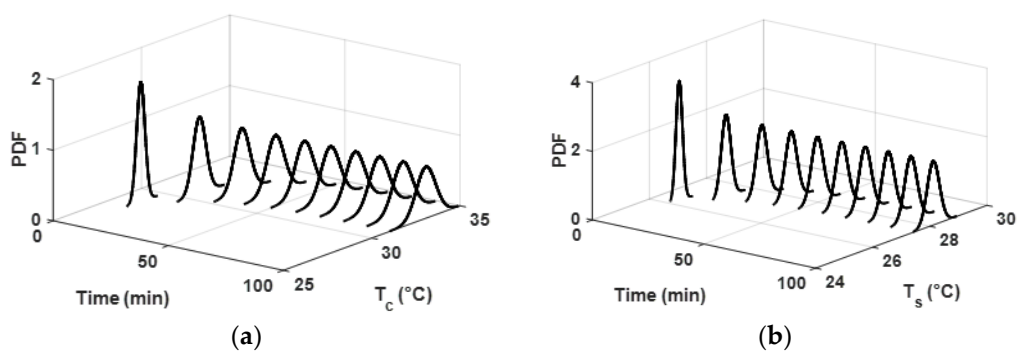

**Figure 10.** Distribution of temperatures for a few time intervals estimated with the gPC coefficients and the correction gains, which can be used to define a two-dimensional domain to generate JCR profiles for FDD: (a) Core temperature approximated with gPC and correction gains and (b) Surface temperature approximated with gPC and correction gains.

As seen in Table 3, the fault classification rate $r_{FCR}$ can be improved approximately by 25% on average with the correction algorithm defined in Equation (17). In addition, study was conducted to investigate the effect of measurement noise on the accuracy of FDD, and Table 4 shows the results of $r_{FCR}$ with respect to different levels of measurement noise. It can be seen that the measurement noise can significantly affect the accuracy of FDD. For instance, the fault classification rate $r_{FCR}$ is about 73% with a 5% measurement noise in the surface temperatures, which has been decreased about 22%, as compared with the case where the measurement noise is 1%.

**Table 3.** Faults Classification Rate with different JCR Profiles.

| $r_{FCR}$ (%) | JCR 1 | JCR 2 | JCR 3 | JCR 4 |
| --- | --- | --- | --- | --- |
| without correction | 59.1 | 62.3 | 59.9 | 69.7 |
| with correction | 89.6 | 89.7 | 82.7 | 88.4 |

**Table 4.** Faults Classification Rate of the Model Corrected by Optimization-based Model Correction.

| | 1% | 2% | 3% | 4% | 5% |
| --- | --- | --- | --- | --- | --- |
| $r_{FCR}$ (%) | 95 | 89.6 | 84.5 | 78.2 | 72.7 |

Using the gPC models, it was found that the optimization of Equation (17) for one function evaluation can be completed in ~1 second on average and the optimum can be achieved in about 30 iterations, which results in an overall simulation time about ~30 seconds. On the other hand, it was found that if Monte Carlo simulations were used for updating the correction gains with 100 samples, ~5 minutes were required for one evaluation of the optimization in Equation (17). Thus, 30 iterations would take ~2.5 hours. This is significantly higher than the gPC-based FDD method, which may be computationally prohibitive for a real-time application of model correction with MC.

## 5. Conclusions

Lithium-ion (Li-ion) batteries are widely used due to its higher energy density and longer life as compared to other batteries. However, the thermal behavior can greatly affect the safety, durability, and performance of Li-ion batteries. Fault detection and diagnosis (FDD), as a key component of the battery management system, play an important role in the management of Li-ion batteries. This paper presents a stochastic FDD algorithm to identify thermal dynamic faults such as thermal runaway fault in Li-ion battery using generalized polynomial chaos (gPC) expansion models. The proposed algorithm consists of three consecutive procedures: (*i*) uncertainty propagation with gPC models to evaluate the effect of uncertainty on measured quantities, which can be used for FDD; (*ii*) accurate fault diagnosis with JCR profiles, which can provide the probabilistic information of being in a faulty operating mode; (*iii*) recursive optimization to adjust the FDD algorithm to account for mismatch between models and thermal dynamics of Li-ion battery cells. It was found that the gPC-based FDD method can outperform sampling-based techniques such as Monte Carlo (MC) simulations in terms of computational efficiency and FDD accuracy. This ensures its on-line applications in Li-ion battery systems such as electric and hybrid electric vehicles. However, the application of the proposed FDD algorithm in complex systems is not purposed for brevity and left for future study. In addition, it is assumed that the uncertainty in this work follows standard distribution in the Askey-Wiener scheme for algorithm clarity. For other distributions, the arbitrary gPC algorithm as explained in our previous work can be used to improve the computational efficiency [40].

**Appendix A. Results of the gPC expansion for the lumped thermal model of Li-ion battery**

$$\frac{dT_{c0}}{dt} = \frac{1}{C_c}\left(I_0^2 R_e + I_1^2 R_e + \frac{1}{R_{c0}}\left((T_{s0}-T_{c0})A + (T_{s2}-T_{c2})B + (T_{s4}-T_{c4})C\right)\right) \quad (A1)$$

$$\frac{dT_{s0}}{dt} = \frac{1}{C_s}\left(\frac{1}{R_u}(T_f - T_{s0}) - \frac{1}{R_{c0}}\left((T_{s0}-T_{c0})A + (T_{s2}-T_{c2})B + (T_{s4}-T_{c4})C\right)\right) \quad (A2)$$

$$\frac{dT_{c1}}{dt} = \frac{1}{C_c}\left(2I_0 I_1 R_e + \frac{1}{R_{c0}}\left((T_{s1}-T_{c1})A + (T_{s5}-T_{c5})B\right)\right) \quad (A3)$$

$$\frac{dT_{s1}}{dt} = \frac{1}{C_s}\left(\frac{1}{R_u}(-T_{s1}) - \frac{1}{R_{c0}}\left((T_{s1}-T_{c1})A + (T_{s5}-T_{c5})B\right)\right) \quad (A4)$$

$$\frac{dT_{c2}}{dt} = \frac{1}{C_c}\left(\frac{1}{R_{c0}}\left((T_{s0}-T_{c0})B + (T_{s2}-T_{c2})C + (T_{s4}-T_{c4})D\right)\right) \quad (A5)$$

$$\frac{dT_{s2}}{dt} = \frac{1}{C_s}\left(\frac{1}{R_u}(-T_{s2}) - \frac{1}{R_{c0}}\left((T_{s0}-T_{c0})B + (T_{s2}-T_{c2})C + (T_{s4}-T_{c4})D\right)\right) \quad (A6)$$

$$\frac{dT_{c3}}{dt} = \frac{1}{C_c}\left(I_1^2 R_e E + \frac{1}{R_{c0}}\left((T_{s3}-T_{c3})FA\right)\right) \quad (A7)$$

$$\frac{dT_{s3}}{dt} = \frac{1}{C_s}\left(\frac{1}{R_u}(-T_{s3})F - \frac{1}{R_{c0}}\left((T_{s3}-T_{c3})FA\right)\right) \quad (A8)$$

$$\frac{dT_{c4}}{dt} = \frac{1}{C_c}\left(\frac{1}{R_{c0}}\left((T_{s0}-T_{c0})C + (T_{s2}-T_{c2})D + (T_{s4}-T_{c4})G\right)\right) \quad (A9)$$

$$\frac{dT_{s4}}{dt} = \frac{1}{C_s}\left(\frac{1}{R_u}(-T_{s4})H - \frac{1}{R_{c0}}\left((T_{s0}-T_{c0})C + (T_{s2}-T_{c2})D + (T_{s4}-T_{c4})G\right)\right) \quad (A10)$$

$$\frac{dT_{c5}}{dt} = \frac{1}{C_c}\left(\frac{1}{R_{c0}}\left((T_{s1}-T_{c1})B + (T_{s5}-T_{c5})C\right)\right) \quad (A11)$$

$$\frac{dT_{s5}}{dt} = \frac{1}{C_s}\left(\frac{1}{R_u}(-T_{s5}) - \frac{1}{R_{c0}}\left((T_{s1}-T_{c1})B + (T_{s5}-T_{c5})C\right)\right) \quad (A12)$$

where *A, B, C, D, E, F, G,* and *H* are all constants calculated with the Galerkin Projection.

## Appendix B Definition and description of faults and their mean values

**Table B1.** Faults Definition and Description.

| JCRs (Mode) | Mean values | Type |
|---|---|---|
| JCR 1 (Faulty 1) | $I = 16.2$, $R_c = 1.68$ | Individual fault |
| JCR 2 (Faulty 3) | $I = 16.2$, $R_c = 2.28$ | Simultaneous faults |
| JCR 3 (Normal) | $I = 13.8$ $R_c = 1.68$ | No fault |
| JCR 4 (Faulty 2) | $I = 13.8$ $R_c = 2.28$ | Individual fault |

## Appendix C Summary of comparison between gPC and MC

**Table C1.** Comparison results between gPC and MC.

| Method | Classification rate | Computational time |
|---|---|---|
| gPC | 0.94 | 5 s |
| MC (100 samples) | 0.75 | 324 s* |

*\* Per optimization iteration of Equation (17)*